# Cubic ReSTe as a High-Performance Thermoelectric Material

Haruka Matsumoto[1], Hiroto Isomura[1], Keita Kojima[1], Ryutaro Okuma[1], Hironori Ohshima[2], Chul-Ho Lee[2], Youichi Yamakawa[3], and Yoshihiko Okamoto[1,*]

[1]Institute for Solid State Physics, the University of Tokyo, Kashiwanoha 5-1-5, Kashiwa 277-8581, Japan.
[2]National Institute of Advanced Industrial Science and Technology (AIST), Tsukuba 305-8568, Japan.
[3]Department of Physics, Nagoya University, Furo-cho, Chikusa-ku, Nagoya 464-8602, Japan.
a) Author to whom correspondence should be addressed: yokamoto@issp.u-tokyo.ac.jp

We report thermoelectric properties of sintered samples of undoped, W-doped, and Sb-doped ReSTe crystallized in a cubic MoSBr-type structure. All samples exhibited $p$-type thermoelectric properties. ReSTe and $Re_{0.993}W_{0.007}STe$ exhibited the largest dimensionless figure of merit $ZT$, reaching 0.4 at 660 K. This high performance is attributed to large power factor owing to the degenerate semiconducting state realized by the strong spin–orbit coupling and low lattice thermal conductivity of the sintered samples. Furthermore, electronic band dispersion of ReSTe is almost flat at the bottom of the conduction band, suggesting that $n$-type ReSTe is expected to exhibit much higher performance than $p$-type ReSTe.

Tellurides are one of the most suitable materials for thermoelectricity. $Bi_2Te_3$-based materials are widely used in commercial Peltier devices, and PbTe and $(AgSbTe_2)_{1-x}(GeTe)_x$ (TAGS) have been used in radioisotope thermoelectric generators [1, 2]. Since tellurides are often narrow-gap semiconductors, the conduction carrier density can be easily controlled to an appropriate value for thermoelectric materials (typically ~$10^{19}$ cm$^{-3}$). Furthermore, spatially extended $5p$ orbitals of tellurium provide high mobility for conduction carriers, and the heavy element nature of tellurium suppresses the lattice thermal conductivity $\kappa_{lat}$ to low values. Consequently, tellurides often exhibit a high dimensionless figure of merit $ZT = S^2T/\rho\kappa$, where $S$, $\rho$, $\kappa$, and $T$ represent the Seebeck coefficient, electrical resistivity, thermal conductivity, and temperature, respectively [3–7]. Another promising feature of tellurides is their ability to form a wide variety of crystal structures owing to the various chemical bonds of tellurium atoms. For example, $CsBi_4Te_6$ [8,9], $ZrTe_5$ [10], and $Ta_4SiTe_4$ [11,12], each of which has a one-dimensional crystal structure, exhibit high thermoelectric performance at low temperatures owing to their one-dimensional characteristics.

Cubic materials with complex crystal structures are also promising for thermoelectric conversion. The multivalley electronic structure stemming from the cubic symmetry increases $S$, and the complex crystal structure reduces $\kappa_{lat}$. When atoms are enclosed in highly symmetric and oversized cages in their crystal structures, the rattling effect can further suppress $\kappa_{lat}$ while maintaining a high electrical conductivity. In fact, there are several material families with cubic and complex crystal structures that exhibit high thermoelectric performance, such as filled skutterudite [13,14], clathrate [15,16], and tetrahedrite [17,18]. However, their practical use as thermoelectric materials has not yet been achieved. If the potential of cubic and complex crystal structures is fully exploited, a next-generation practical material with a sufficiently high thermoelectric performance can be achieved. The use of tellurides is a promising strategy for this purpose.

In this paper, we report the thermoelectric properties of ReSTe with cubic and complex crystal structure. ReSTe was first synthesized by Fedorov *et al.* in powder form and was reported to crystallize in a cubic MoSBr type with the space group $F$–$43m$ [20,21]. As shown in Fig. 1(a), the $Re_4S_4$ cubanoclusters surrounded by Te atoms are arranged to form a face-centered cubic structure. The Re atoms are coordinated by $S_3Te_3$ octahedra to form a breathing pyrochlore structure [22]. The electronic states of the $Re_4S_4$ clusters were studied by X-ray emission and photoelectron spectroscopy [23]. However, their transport properties have not been reported. We synthesized undoped, W-doped, and Sb-doped ReSTe sintered samples and investigated their thermoelectric properties. The undoped and 0.7% W-doped samples exhibited high $p$-type thermoelectric performances, with $ZT =$ 0.4 at 660 K. According to the first-principles calculations, ReSTe exhibits a much smaller band gap at the Fermi level than layered Re dichalcogenides and an almost flat band at the bottom of the conduction band. ReSTe contains rhenium, which is one of the rarest elements on earth, and decomposes at 700–750 K, suggesting that this material is not suitable for practical use. However, the results obtained in this study demonstrate the high potential of tellurides with cubic and complex crystal structures for thermoelectric applications.

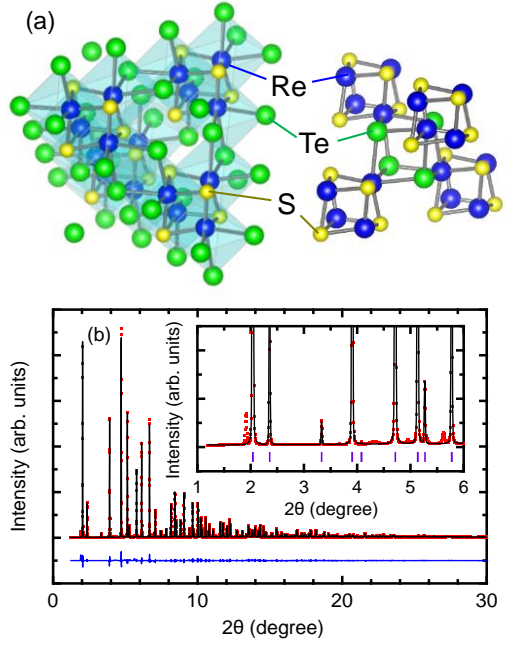

Figure 1. (a) Crystal structure of ReSTe [19]. (b) A synchrotron powder XRD pattern of the sintered ReSTe sample taken at 300 K. The X-ray wavelength was $\lambda = 0.20675$ Å. Solid circles and overplotted curves represent the experimental data and calculated pattern, respectively. The lower curve in the main panel shows a plot difference between the experimental and calculated intensities. The inset shows an enlarged view of the low-angle region. The vertical bars indicate the position of the Bragg reflections of ReSTe. The reliability factors of the refinement are $wR_p = 8.34\%$, $R_p = 4.89\%$, and $S = 1.50$.

Sintered samples of ReSTe, $Re_{1-x}W_xSTe$ ($x \leq 0.05$), and $ReSTe_{1-y}Sb_y$ ($y \leq 0.05$) were synthesized using a conventional solid-state reaction method. Stoichiometric amounts of Re, W, S, Te, and Sb powders were mixed, pressed into pellets, and sealed in evacuated quartz tubes. The tubes were heated and maintained at 673 K for 24 h and then at 1273 K for 48 h (1073 K for $y = 0.05$) with intermediate grinding. Powder X-ray diffraction (XRD) measurements were performed at BL13XU in SPring-8 using a Debye–Scherrer diffractometer with a 60 keV X-ray. Figure 1(b) shows the result of Rietveld analysis of a powder XRD pattern of ReSTe using JANA2006 [24]. In this XRD pattern, where all the diffraction peaks, except for small peaks due to the $ReS_2$ and Te impurity phases, are indexed on the basis of a cubic unit cell with the $F\bar{4}3m$ space group, indicating that cubic ReSTe was obtained as an almost single phase. As mentioned in Supplementary Note 1, the W- or Sb-doped ReSTe samples had the same quality as the undoped sample. Each sample prepared in this study exhibited a nominal chemical composition.

The electrical resistivity, Seebeck coefficient, and thermal conductivity below 350 K were measured using the continuous measurement mode of the thermal transport option of a Physical Property Measurement System (PPMS, Quantum Design). The Hall resistivity below 300 K was measured using the PPMS. The electrical resistivity and Seebeck coefficient of the undoped and $x = 0.007$ samples above room temperature were measured using the dc four-probe and quasi-steady temperature differential methods, respectively, with a ZEM3 (ADVANCE RIKO) under a He atmosphere. The thermal diffusivity of the undoped and $x = 0.007$ samples was measured using the laser flash method with an LFA457 (Netzsch) under an Ar gas flow. The thermal conductivity was calculated as $\kappa = DC_p d_s$, where $D$, $C_p$, and $d_s$ denote the thermal diffusivity, heat capacity, and sample density, respectively. First-principles calculations for ReSTe were performed using the WIEN2k code within the density functional theory (DFT) framework, employing the full-potential linearized augmented plane wave (FP-LAPW) method [25]. The generalized gradient approximation (GGA) in the Perdew–Burke–Ernzerhof (PBE) form was adopted for the exchange-correlation functional. A $k$-point mesh of $20\times20\times20$ was used to sample the Brillouin zone, and the energy-convergence criterion was set to 0.0001 Ry. The experimental structural parameters were used in the calculations.

Figure 2 shows the temperature dependences of the $S$, $\rho$, and $\kappa$ of the sintered ReSTe, $Re_{1-x}W_xSTe$ ($x \leq 0.05$), and $ReSTe_{1-y}Sb_y$ ($y \leq 0.05$) samples. As shown in Fig. 2(a), the undoped ReSTe exhibited a positive $S$ increasing with increasing temperature with a concave downward curve. The value of $S$ was the largest among all the samples, exceeding 260 µV K$^{-1}$ at 300 K, which is high enough to be a thermoelectric material. The $\rho$ of undoped ReSTe exhibits a semiconducting temperature dependence, with $\rho$ exponentially increasing with decreasing temperature. The value of $\rho = 30$ mΩ cm at 300 K is one order of magnitude higher than that of practical materials. With increasing W content $x$, both $S$ and $\rho$ systematically decrease, and the temperature dependence of $\rho$ becomes weaker. These results indicate that the substituted W atoms act as hole dopant, which is consistent with the hole-carrier concentration estimated from the Hall resistivity data (discussed later). In the case of $ReSTe_{1-y}Sb_y$, in which Sb atoms substituted the Te sites, $S$ was systematically reduced, as in the case of $Re_{1-x}W_xSTe$; however, the decrease in $\rho$ was weaker. These results indicate that Sb doping also acts as hole doping and perhaps has a greater disorder effect on the transport properties than W doping, suggesting a higher appropriateness of W doping for optimizing the thermoelectric properties of ReSTe. The thermoelectric power factor, $P = S^2/\rho$, at 300 K reached a maximum value of 3.1 µW cm$^{-1}$ K$^{-2}$ for $x = 0.007$ and 0.01. Although large, this value was one order of magnitude smaller than the practical level.

As shown in Fig. 2(c), the undoped sample exhibits a low $\kappa$ of 17 mW cm$^{-1}$ K$^{-1}$ at 300 K. The value of $\kappa$ at 300 K increases with increasing $x$, mainly owing to the increase of the electron contribution of thermal conductivity $\kappa_{el}$. Figure 3(a) shows the lattice contribution of thermal conductivity, $\kappa_{lat} = \kappa - \kappa_{el}$, where $\kappa_{el}$ is estimated using the Wiedemann–Franz law of $\kappa_{el}\rho/T = L = 2.44 \times 10^{-8}$ V$^2$ K$^{-2}$. All W-doped samples exhibit almost the same $\kappa_{lat}$ of approximately 16 mW cm$^{-1}$ K$^{-1}$, which is small for a sintered sample of a crystalline material. The low $\kappa_{lat}$ can be attributed to the content of heavy Re and Te atoms in ReSTe. The structural parameters determined from the synchrotron XRD data, as shown in Table I, indicated no atomic sites with unusually large atomic displacement parameters $U_{iso}$, in contrast to clathrate and filled skutterudite. Future studies should focus on determining the origin of the low $\kappa_{lat}$ in ReSTe and investigate the contribution of the complex crystal structure of ReSTe to it. In the Sb-doped samples, $\kappa_{lat}$ decreased further with increasing $y$, probably owing to the disorder induced by Sb substitution, as for $\rho$.

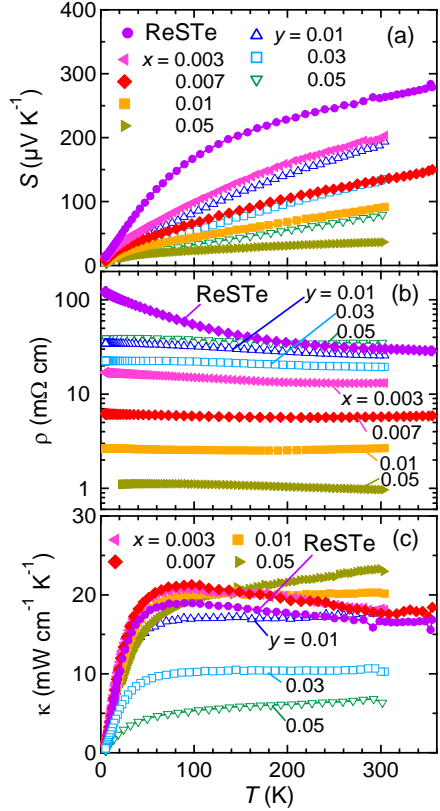

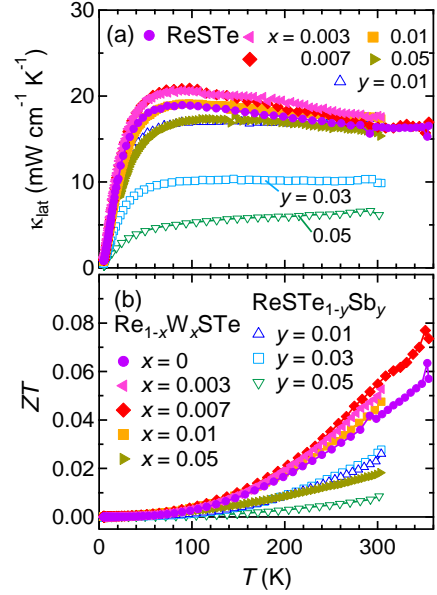

Figure 2. Temperature dependences of the (a) Seebeck coefficient, (b) electrical resistivity, and (c) thermal conductivity of the sintered $Re_{1-x}W_x STe$ ($x \leq 0.05$) and $ReSTe_ySb_{1-y}$ ($y \leq 0.05$) samples measured below 350 K.

Figure 3. Temperature dependences of the (a) lattice thermal conductivity $\kappa_{lat}$ and (b) dimensionless figure of merit $ZT$ of the sintered $Re_{1-x}W_x STe$ ($x \leq 0.05$) and $ReSTe_ySb_{1-y}$ ($y \leq 0.05$) samples.

Table I. Crystallographic parameters for ReSTe determined by the Rietveld analysis of synchrotron powder XRD data shown in Fig. 1(b). The space group is $F\bar{4}3m$. The lattice parameter and unit cell volume are $a = 10.5217(2)$ Å and $V = 1015.732(4)$ Å$^3$, respectively. The occupancy of each site was fixed at one because refinements converged in a model in which each atom fully occupied each Re, S, and Te site.

|    |     | $x$        | $y$ | $z$ | $U_{iso}$ (Å$^2$) |
|----|-----|------------|-----|-----|-------------------|
| Re | 16d | 0.34817(2) | $x$ | $x$ | 0.00335(5)        |
| S  | 16d | 0.11795(14)| $x$ | $x$ | 0.0037(4)         |
| Te | 16d | 0.62300(4) | $x$ | $x$ | 0.00463(9)        |

As shown in Fig. 3(b), the $ZT$ values evaluated using these physical properties increased with increasing temperature for all the samples. Over the entire temperature range below room temperature, the lightly W-doped samples showed a large $ZT$; the $x = 0.007$ sample exhibited the largest $ZT = 0.055$ at 300 K, whereas the undoped sample exhibited $ZT = 0.04$ at the same temperature. Although these $ZT$ values are much smaller than the practical values, a larger $ZT$ is expected at higher temperatures.

Figure 4 shows the thermoelectric properties measured above room temperature for the undoped and $x = 0.007$ sintered samples. The measured data for the heating and cooling processes were equal, indicating that the samples did not degrade within the measured temperature range. As seen from Fig. 4(c), the $\kappa_{lat}$ for both samples weakly decreased with increasing temperature. In the $x = 0.007$ sample, $S$ and $\rho$ were almost constant, as shown in Figs. 4(a) and 4(b). In contrast, the $S$ and $\rho$ of the undoped sample significantly decreased above 400 K, probably owing to the thermal excitations across the band gap. For $x = 0.007$, this effect was suppressed by hole doping, resulting in the almost constant $S$ and $\rho$. As shown in Fig. 4(d), the $ZT$ values of the undoped and $x = 0.007$ samples increased significantly with increasing temperature. They exhibit $ZT = 0.39$ and 0.36, respectively, at the highest measured temperature of 660 K.

These samples were expected to exhibit higher $ZT$ at higher temperatures. However, this was difficult to achieve because ReSTe decomposed between 700 and 750 K. This result and the fact that rhenium is one of the rarest elements suggest that ReSTe is not suitable for practical use. Nevertheless, the high performance of ReSTe ($ZT$ = 0.4 at 660 K) demonstrates the high potential of tellurides with cubic and complex crystal structures for thermoelectric applications.

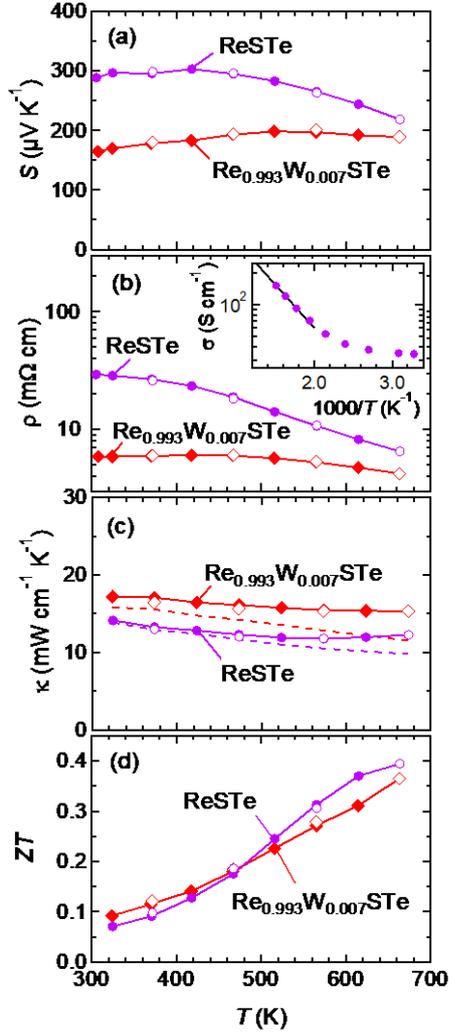

Figure 4. Temperature dependences of the (a) Seebeck coefficient, (b) electrical resistivity, (c) thermal conductivity, and (d) dimensionless figure of merit $ZT$ of the sintered ReSTe and $Re_{0.993}W_{0.007}STe$ samples. The data in the heating (filled) and cooling (open) processes were measured in the given order. The broken lines in (c) show the lattice thermal conductivity $\kappa_{lat} = \kappa - \kappa_{el}$, where $\kappa_{el}$ is estimated using the Wiedemann–Franz law of $\kappa_{el}\rho/T = L = 2.44 \times 10^{-8}$ V$^2$ K$^{-2}$. The inset in (b) shows an Arrhenius plot of electrical conductivity of ReSTe. The solid line represents the result of a linear fit to the data above 550 K.

The large $P$ realized in lightly W-doped ReSTe, which is an important factor for the high thermoelectric performance of this system, is discussed based on the Hall resistivity and first-principles calculation data. The Hall resistivity of the $x$ = 0.01 sintered sample increased linearly with the application of magnetic fields at all measured temperatures below 300 K. This indicated a dominance of the hole carriers, consistent with the $S$ data. The $n$ was almost independent of temperature with $n = 4.6 \times 10^{19}$ cm$^{-3}$ at 300 K. Hole carriers were also dominant in the undoped sample, with $n = 1.8 \times 10^{18}$ cm$^{-3}$ at 300 K, which was significantly smaller than that for $x$ = 0.01. This result indicates that the W doping acts as hole doping and the large $P$ is realized at $n \sim 10^{19}$ cm$^{-3}$, comparable to those of the practical materials. The $\mu$ of $x$ = 0.01 and undoped samples were 76 and 83 cm$^2$ V$^{-1}$ s$^{-1}$ at 300 K, respectively, which are lower than those of Bi$_2$Te$_3$ and PbTe [1]. This is likely due to not only the material's intrinsic properties, such as the band width, but also to extrinsic factors such as sinterability of the sample.

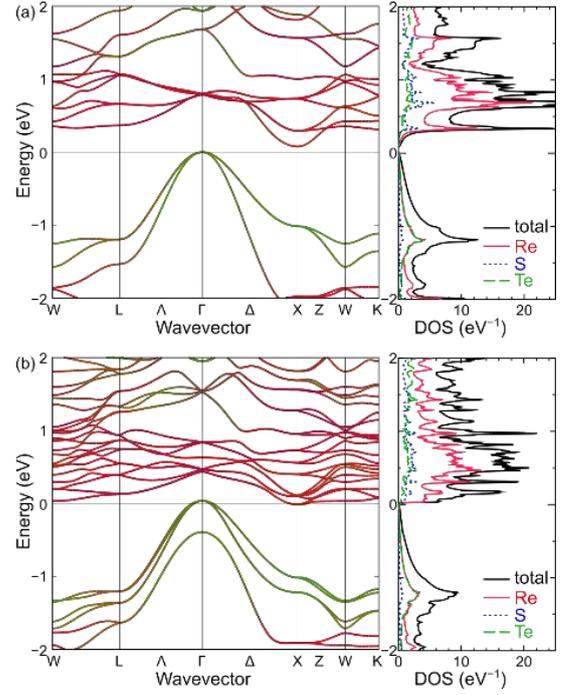

Figure 5. Electronic structures (left) and density of states (right) of ReSTe (a) without and (b) with spin–orbit coupling. The red, blue, and green colors in the electronic structures represent the contributions of Re, S, and Te, respectively. The Fermi level is set to 0 eV.

The electronic components affecting the thermoelectric properties of ReSTe are discussed based on its electronic structure. Figure 5(a) shows the first-principles calculation results for ReSTe calculated without spin–orbit coupling. A small energy gap of approximately 0.1 eV was observed between the valence and conduction bands. By switching on the spin–orbit coupling, as shown in Fig. 5(b), the band dispersions at the top of the valence band located at the Γ point split by approximately 0.4 eV, pushing the top of the valence band to a higher energy. This caused the top of the valence band to overlap with the bottom of the conduction band, resulting in a semimetallic band structure. However, considering the underestimation of the band gap in the local-density approximation calculations, a small band gap most likely exists instead of such a "spin–orbit coupling induced semimetallic state" in reality. The presence of a small band gap is consistent with the observed ρ data. A linear fit to the Arrhenius plot of electrical conductivity of undoped ReSTe above 550 K, as shown in the inset of Fig. 4(b), yields the activation energy of $E_a$ = 0.16 eV, suggesting a small energy gap corresponding to the $E_a$ opening at the Fermi level. This band gap was considerably smaller than that of layered Re dichalcogenides with the same electron filling as that of ReSTe. $ReS_2$ and $ReTe_2$ are band insulators with an energy gap of more than 1 eV [26–28]. Materials with a small band gap are suitable as thermoelectric materials because low ρ and large S can be realized by controlling n through chemical doping. In the W-doped samples, the acceptor levels introduced by the W doping were most likely formed near the top of the valence band, resulting in a p-type degenerate semiconducting state that exhibits high thermoelectric performance.

By comparing Figs. 5(a) and 5(b), the dispersion at the bottom of the conduction band is found to become flat upon switching on the spin–orbit coupling. This is probably due to the stronger hybridization between the conduction band and the top of the valence band caused by the band splitting. Consequently, density of states at the bottom of the conduction band becomes much steeper, as shown in Fig. 5(b), suggesting that n-type ReSTe can achieve a much higher thermoelectric performance than that in p-type. Given that a fairly high ZT of 0.4 (T = 660 K) has already been realized in p-type samples, the ZT of the n-type samples may exceed the practical level. Unfortunately, we have not succeeded in synthesizing n-type ReSTe, despite using various chemical doping methods, such as Ru-, Os-, and Ir-doping of the Re sites and I-doping of the Te sites. However, an n-type ReSTe should be synthesized in future studies.

In summary, sintered samples of undoped and chemically-doped ReSTe were synthesized, and their thermoelectric properties were evaluated. The results showed that the undoped and 0.7% W-doped (x = 0.007) samples exhibited a high p-type thermoelectric performance at high temperatures, reaching ZT = 0.4 at 660 K. The large P due to the degenerate semiconducting state and the low $\kappa_{lat}$ play important roles in achieving the high thermoelectric performance in ReSTe. Furthermore, the electronic band dispersion at the bottom of the conduction band is almost flat owing to the large band splitting in the valence band caused by the strong spin–orbit coupling. This result strongly suggests that a high thermoelectric performance far beyond that of the p-type can be realized in n-type samples, indicating the great potential of tellurides with cubic and complex crystal structures as thermoelectric materials.

**Acknowledgments**
The authors are grateful to H. Nishiate for advice on the thermoelectric property measurements. This study was supported by JSPS KAKENHI (Grant Nos. 23H01831 and 23K26524), JST ASPIRE (Grant No. JPMJAP2314), and JST Mirai (Grant No. JPMJMI19A1). The powder XRD experiments were conducted at SPring-8 (Proposal Nos. 2024A1528 and 2024B2007) in Hyogo, Japan.

**References**
[1] G. D. Mahan, Solid State Physics (Academic Press, New York, USA, 1997) Vol. 51, pp. 81-157.
[2] R. D. Abelson, Thermoelectrics Handbook Macro to Nano (Taylor & Francis Group, Boca Raton, USA, 2006) pp. 56-1-29.
[3] K. F. Hsu, S. Loo, F. Guo, W. Chen, J. S. Dyck, C. Uher, T. Hogan, E. K. Polychroniadis, and M. G. Kanatzidis, Science **303**, 818-821 (2004).
[4] R. F. Brebrick, J. Phys. Chem. Solids **24**, 27-36 (1963).
[5] R. F. Brebrick and A. J. Strauss, Phys. Rev. **131**, 104 (1963).
[6] T. Plirdpring, K. Kurosaki, A. Kosuga, T. Day, S. Firdosy, V. Ravi, G. J. Snyder, A. Harnwunggmoung, T. Sugahara, Y. Ohishi, H. Muta, and S. Yamanaka, Adv. Mater. **24**, 3622 (2012).
[7] Y. Luo, J. Yang, Q. Jiang, W. Li, D. Zhang, Z. Zhou, Y. Cheng, Y. Ren, and X. He, Adv. Energy Mater. **6**, 1600007 (2016).
[8] D.-Y. Chung, T. Hogan, P. Brazis, M. Rocci-Lane, C. Kannewurf, M. Bastea, C. Uher, and M. G. Kanatzidis, Science **287**, 1024 (2000).
[9] D.-Y. Chung, T. P. Hogan, M. Rocci-Lane, P. Brazis, J. R. Ireland, C. R. Kannewurf, M. Bastea, C. Uher, and M. G. Kanatzidis, J. Am. Chem. Soc. **126**, 6414 (2004).
[10] R. T. Littleton IV, T. M. Tritt, J. W. Kolis, D. R. Ketchum, N. D. Lowhorn, and M. B. Korzenski, Phys. Rev. B **64**, 121104 (2001).
[11] T. Inohara, Y. Okamoto, Y. Yamakawa, A. Yamakage, and K. Takenaka, Appl. Phys. Lett. **110**, 183901 (2017).
[12] Y. Okamoto, Y. Yoshikawa, T. Wada, and K. Takenaka, Appl. Phys. Lett. **115**, 043901 (2019).
[13] B. C. Sales, D. Mandrus, and R. K. Williams, Science **272**, 1325-1328 (1996).
[14] X. Shi, J. Yang, J. R. Salvador, M. Chi, J. Y. Cho, H. Wang, S. Bai, J. Yang, W. Zhang, L. Chen, J. Am. Chem. Soc. **133**, 7837-7846 (2011).
[15] G. S. Nolas, J. L. Cohn, G. A. Slack, S. B. Schujman, Appl. Phys. Lett. **73**, 178-180 (1998).
[16] B. C. Sales, B. C. Chakoumakos, R. Jin, J. R. Thompson, and D. Mandrus, Phys. Rev. B **63**, 245113 (2001).
[17] K. Suekuni, K. Tsuruta, M. Kunii, H. Nishiate, E. Nishibori, S. Maki, M. Ohta, A. Yamamoto, and M. Koyano, J. Appl. Phys. **113**, 043712 (2013).
[18] X. Lu, D. T. Morelli, Y. Xia, F. Zhou, V. Ozolins, H. Chi, X. Zhou, and C. Uher, Adv. Energy Mater. **3**, 342-348 (2013).
[19] K. Momma and F. Izumi, J. Appl. Cryst. **44**, 1272 (2011).
[20] V. E. Fedorov, Y. V. Mironov, V. P. Fedin, and Y. I. Mironov. J. Struct. Chem. **35**, 146 (1994).

[21] V. E. Fedorov, Y. V. Mironov, V. P. Fedin, H. Imoto and T. Saito, Acta Cryst. **C52**, 1065-1067 (1996).

[22] Y. Okamoto, G. J. Nilsen, J. P. Attfield, and Z. Hiroi, Phys. Rev. Lett. **110**, 097203 (2013).

[23] G. F. Khudorozhko, É. A. Kravtsova, L. N. Mazalov, V. E. Fedorov, L. G. Bulusheva, I. P. Asanov, G. K. Parygina, and Y. V. Minorov, J. Struct. Chem. **37**, 767 (1996).

[24] V. Petříček, M. Dušek, and L. Palatinus, Z. Kristallogr. Cryst. Mater. **229**, 345 (2014).

[25] P. Blaha, K. Schwarz, G. Madsen, D. Kvasnicka, and J. Luitz: WIEN2k, An Augmented Plane Wave + Local Orbitals Program for Calculating Crystal Properties (Techn. Universität Wien, Austria, 2001).

[26] K. K. TIong, C. H. Ho, and Y. S. Huang, Solid State Commun. **111**, 635 (1999).

[27] S. Tongay, H. Sahin, C. Ko, A. Luce, W. Fan, K. Liu, J. Zhou, Y.-S. Huang, C.-H. Ho, J. Yan, D. F. Ogletree, S. Aloni, J. Ji, S. Li, J. Li, F. M. Peeters, and J. Wu, Nat. Commun. **5**, 3252 (2014).

[28] D. Banik, C. Lionel, S. Das, and S. Koley, Phys. Scr. **99**, 085975 (2024).

**Supplementary Note 1. Powder X-ray diffraction patterns of chemically-doped ReSTe**

Figure S1 shows the powder X-ray diffraction (XRD) patterns of the sintered samples of undoped ReSTe, Re$_{1-x}$W$_x$STe ($x \leq 0.05$), and ReSTe$_{1-y}$Sb$_y$ ($y \leq 0.05$). The W- and Sb-doped samples showed a diffraction pattern similar to that of the undoped sample. All the diffraction peaks, except for the small peaks from the ReS$_2$ and unknown impurity phases, were indexed on the basis of cubic $F\bar{4}3m$ symmetry, indicating that almost single-phase samples of W- or Sb-doped ReSTe were obtained.

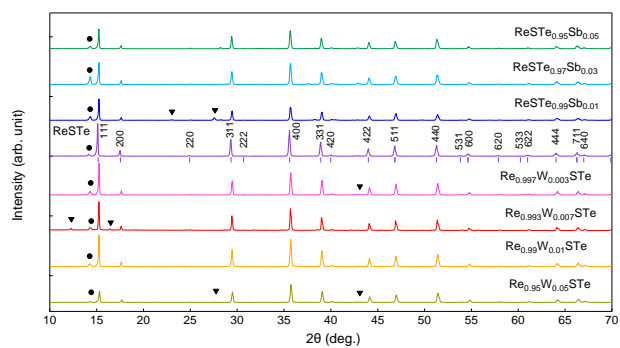

Supplementary Figure 1. Powder XRD patterns of sintered samples of undoped ReSTe, Re$_{1-x}$W$_x$STe ($x \leq 0.05$), and ReSTe$_{1-y}$Sb$_y$ ($y \leq 0.05$) measured at room temperature. Filled circles and triangles indicate the diffraction peaks of ReS$_2$ and unknown impurities, respectively.